# Compound-Specific Chlorine Isotope Analysis of Organochlorines Using Gas Chromatography-Double Focus Magnetic-Sector High Resolution Mass Spectrometry

Caiming Tang[1,2,*] and Jianhua Tan[3]

[1] State Key Laboratory of Organic Geochemistry, Guangzhou Institute of Geochemistry, Chinese Academy of Sciences, Guangzhou 510640, China

[2] University of Chinese Academy of Sciences, Beijing 100049, China

[3] Guangzhou Quality Supervision and Testing Institute, Guangzhou 510110, China

*Supporting Information.*

**ABSTRACT:** Compound-specific chlorine isotope analysis (CSIA-Cl) is a practicable and high-performance approach for quantification of transformation processes and pollution source apportionment of chlorinated organic compounds. This study developed a CSIA-Cl method for perchlorethylene (PCE) and trichloroethylene (TCE) using gas chromatography-double focus magnetic-sector high resolution mass spectrometry (GC-DFS-HRMS) with a bracketing injection mode. The achieved highest precision for PCE was 0.21‰ (standard deviation of isotope ratios), and that for TCE was 0.25‰. When one standard was used as the external isotopic standard for another of the same analyte, the lowest standard deviations of relative isotope-ratio variations ($\delta^{37}Cl'$) between the two corresponding standards were 0.64‰ and 0.80‰ for PCE and TCE, respectively. As a result, the critical $\delta^{37}Cl'$ for differentiating two isotope ratios are 2.6‰ and 3.2‰ for PCE and TCE, respectively, which are comparable with those in some reported studies using GC-quadrupole MS (GC-qMS). The lower limit of detection for CSIA-Cl of PCE was 0.1 μg/mL (0.1 ng on column), and that for TCE was determined to be 1.0 μg/mL (1.0 ng on column). Two isotope ratio calculation schemes, i.e., a scheme using complete molecular-ion isotopologues and another one using a pair of neighboring isotopologues, were evaluated in terms of precision and accuracy. The complete-isotopologue scheme showed evidently higher precision and was deduced to be more competent to reflect trueness in comparison with the isotopologue-pair scheme. Influences of isotope fractionation occurring on GC-MS to CSIA-Cl results were assessed, and the isotope fractionation was found to have no significantly negative effect on the precision and trueness of CSIA-Cl if external isotopic standards were utilized. Temporal drifts of isotope ratios with injection sequences were observed for some standards, suggesting the mandatory application of external isotopic standards in CSIA-Cl using GC-HRMS. The CSIA-Cl method developed in this study will be conducive to future studies concerning transformation processes and source apportionment of PCE and TCE, and light the ways to method development of CSIA-Cl for more organochlorines. (**Preprint version submitted to ArXiv.org.**)

## INTRODUCTION

Chlorinated organic compounds such as perchlorethylene (PCE) and trichloroethylene (TCE) are notorious and ubiquitous pollutants in the environment.[1] Compound-specific isotope analysis (CSIA) has become a mature approach in quantification of transformation processes of organic compounds, pollution source apportionment, and doping control.[2] The recently developed dual-element isotope analysis has more powerful features in quantifying attenuation and identifying sources for organic pollutants.[3-6] CSIA of chlorine (CSIA-Cl) has been developed in last decades, and attracting increasing scientific concerns in environmental research fields.[7] While it is still an emerging tool on the way to improvement in environmental analysis. The conventional CSIA-Cl methods use cumbersome offline or online conversion of target organochlorines into compounds possessing only one chlorine atom (e.g., CsCl, $CH_3Cl$ and HCl), which are suitable for the isotope ratio detection using traditional multiple-collector isotope ratio mass spectrometry (IRMS).[8-13] The first CSIA-Cl method using gas chromatography-IRMS (GC-IRMS) without the conversion (combustion/pyrolysis) step was developed by Shouakar-Stash et al. through directly monitoring the select isotopologue ions of chloroethylenes.[9]

CSIA-Cl methods using regular GC-quadrupole mass spectrometry (GC-qMS) have been developed during the last decade.[3,14-18] Recently, GC-quadrupole hybrid time-of-flight (TOF) MS (GC-QTOF-MS) has been employed to perform CSIA-Cl of hexachlorobezene.[19] All these emerging CSIA-Cl methods using GC-qMS or GC-QTOF-MS calculated the isotope ratios using the first pair of two-mass apart chlorine isotopologues of either the molecular ion or a fragmental ion of an analyte, or using the first isotopologue pairs of all the molecular and the fragmental ions. All these studies tacitly approved that the isotopologues of the detected ions of the analytes conformed to binomial distribution, which is the prerequisite assumption when the isotopologue-pair scheme is applied to calculating isotope ratios. A theoretical study has been reported in terms of the calculation of isotope ratios and quantification of isotope fractionation using ion-signal ratios of molecular and/or fragmental ions of polychlorinated compounds.[20] The study stressed that accurate chlorine isotope ratios can theoretically be calculated with any pair of isotopologues of molecular or fragmental ions, in other words, the isotope ratios derived from the molecular ion and the fragmental ions of an analyte should be equivalent. However, chlorine isotope fractionation can occur on GC-MS systems,[21,22] particularly in the EI-MS part. The molecular ion of an organochlorine is anticipated to show less significant isotope fractionation than the product ion(s), because the molecular ion is only influenced by intermolecular isotope fractionation while the product ion(s) may be subjected to both intermolecular and intramolecular isotope fractionations.[23] Thus chlorine isotope ratios calculated with different ions (e.g., molecular and fragmental ions) may vary, if these isotope ratios are not calibrated with proper external isotopic standards. Furthermore, our recent study found that the detected isotopologue ions did not conform to binomial distribution, which has also been proved in theory.[23] In addition, the chlorine isotopologues of synthetic organochlorines with more than one Cl atom produced under some conditions and all environmental polychlorinated organic compounds are unlikely exactly binomial distributed.[24] As a result, the isotope ratio calculation schemes using pairs of neighboring isotopologue ions need reevaluation and reconsideration.

Unlike the isotopologue-pair scheme, the calculation scheme using complete isotopologues can avoid disadvantages caused by non-binomial distributions of detected chlorine isotopologue ions of target organochlorines. In addition, as the molecular-ion isotope fractionation may be less significant than the product-ion isotope fractionation in EI-MS, the uncertainties of isotope ratios derived from molecular ions are deduced to be lower than those derived from fragmental ions. Thus, the complete-isotopologue scheme using molecular-ion isotopologues may be promising for calculating isotope ratios in CSIA-Cl. However, this complete-isotopologue scheme may be inapplicable for GC-qMS when analyzing polychlorinated organic compounds owing more than two chlorine atoms, due to their insufficient signal intensities. While GC-double focus magnetic-sector high resolution MS (GC-DFS-HRMS) can provide significantly higher sensitivity and selectivity for organochlorines than GC-qMS, and thus may be a promising alternative tool for CSIA-Cl.

In this study, we developed a CSIA-Cl method for PCE and TCE using GC-DFS-HRMS in association with the complete-isotopologue scheme of isotope ratio calculation using molecular ions only. The method was partially validated in terms of precision, sensitivity and amount dependency. This method will benefit future studies concerning transformation process quantification and source apportionment of PCE and TCE, and may illuminate the ways to CSIA-Cl method development for more organochlorines.

## EXPERIMENTAL SECTION

### Chemicals and Materials.

Reference standards of perchlorethylene (PCE, 99.0%) and trichloroethylene (TCE, 99.5%) of chromatographic grade were bought from Dr. Ehrenstorfer (Augsburg, Germany), and the analytical-reagent-grade PCE and TCE were obtained from Tianjin Fuyu Chemical Co. Ltd. (Tianjin, China). The standards were accurately weighed and dissolved in n-hexane to prepare stock solutions at 1.0 mg/mL. The stock solutions were further serially diluted with n-hexane to prepare working solutions at two concentration levels (1.0 μg/mL and 0.1 μg/mL). All the standard solutions were stored at -20 °C condition prior to use. Chromatographic-grade n-hexane was purchased from Merck Corp. (Darmstadt, Germany).

### Instrumental Analysis.

The working solutions were directly analyzed by GC-HRMS. The working solutions of the standards stemming from different producers (manufacturer-1 and manufacturer-2) were injected onto the GC-HRMS in a bracketing way as proposed in a previous study,[15] and 5-6 replicated injections were carried out for each working solution in an analysis batch. The GC-HRMS system comprised dual gas chromatographers (Trace-GC-Ultra) coupled with a double focus magnetic-sector high resolution MS and a TriPlus auto-sampler (GC-DFS-HRMS, Thermo-Fisher Scientific, Bremen, Germany). A DB-5MS capillary column (60 m × 0.25 mm, 0.25 μm thickness, J&W Scientific, USA) was used, with helium as the carrier gas (constant flow of 1.0 mL/min). The oven temperature was initially held at 40 °C for 2 min, ramped at 2 °C/min to 65°C, then ramped to 300 °C at 40 °C/min and held for 1 min. The GC inlet and transfer line were set at 260 °C and 280 °C, respectively.

The working conditions and parameters of the HRMS are provided as follows: positive electron ionization source (EI+) was used; EI energy was 45 eV; ionization source was maintained at 250 °C; filament current of EI source was 0.8 mA; multiple ion detection (MID) mode was used for data acquisition; dwell time of each isotopologue ion was 20 ms; mass resolution was ≥ 10000 (5% peak-valley definition) and the HRMS detection accuracy was ± 0.001 u. The MID was started at 7.6 min, and the monitoring time segments were 7.6-11.5 min and 11.5-15 min for TCE and PCE, respectively. The MID cycles were of 120 ms and 140 ms for TCE and PCE, respectively. The HRMS was calibrated in real time with perfluorotributylamine during MID operation.

Structures of the investigated compounds were sketched by ChemDraw (Ultra 7.0, Cambridgesoft), and the exact masses of the molecular isotopologues were calculated with mass accuracy of 0.00001 u. Only chlorine isotopologues were considered. For a compound containing $n$ Cl atoms, the complete isotopologues ($n$+1) were chosen. The mass-to-charge ratios ($m/z$) of the isotopologue ions were obtained by subtracting the mass of an electron from the exact mass of each isotopologue. The $m/z$ values were imported into the MID module for data acquisition. The details related to isotopologues of the investigated compounds, such as retention times, isotopologue chemical formulas, exact masses and exact $m/z$ values are provided in Table S-1.

**Data Processing.**

Chlorine isotope ratio (IR) was calculated as:

$$IR = \frac{\sum_{i=0}^{n} i \times I_i}{\sum_{i=0}^{n} (n-i) \times I_i} \quad (1)$$

where $n$ is the number of Cl atoms of a molecule; $i$ is the number of $^{37}$Cl atoms in an isotopologue ion; $I_i$ is the MS signal intensity of the isotopologue ion $i$.

All the measured isotope ratios in this study were relative values (raw isotope ratios) without being calibrated to standard mean ocean chlorine (SMOC) scale due to unavailability of the external isotopic standards with known chlorine isotope composition and structurally identical to the investigated compounds. Compromisingly, we regarded the standards purchased from different manufacturers (Dr. Ehrenstorfer (manufacturer-1) and Tianjin Fuyu Chemical Co. Ltd. (manufacturer-2)) as the mutual external isotopic standards for each other, for calculation of relative isotope-ratio variations ($\delta^{37}$Cl'):

$$\delta^{37}\text{Cl'} = \left(\frac{IR_{manufacturer-1}}{IR_{manufacturer-2}} - 1\right) \times 1000\text{‰} \quad (2)$$

or

$$\delta^{37}\text{Cl'} = \left(\frac{IR_{manufacturer-2}}{IR_{manufacturer-1}} - 1\right) \times 1000\text{‰} \quad (3)$$

The average MS signal intensity of each isotopologue ion within the whole chromatographic peak was used for isotope ratio calculation. Background subtraction was performed before exporting MS signal intensity by subtracting intensities of the baseline regions adjacent to both ends of the corresponding chromatographic peak. Data from 5-6 replicated injections were applied to evaluate a mean isotope ratio and its standard deviation (1σ).

## RESULTS AND DISCUSSION

**Method Performances.**

*Precision.* Two analysis batches were carried out, of which the first one (batch-I) comprised of 5 replicated injections and the second contained 6 injections for each standard. The standard deviations of the measured isotope ratios were 0.21-0.53‰ and 0.21-0.43‰ in batch-I and batch-II, respectively (Table 1). PCE shown generally higher precisions (standard deviations of isotope ratios: 0.21-0.39‰) than TCE (0.25-0.53‰). The precisions in terms of isotope-ratio standard deviations of the standards from the two different manufacturers were comparable. As for PCE, if the difference between two isotope ratios is larger than 0.0016, the isotope ratios can be confidently differentiated. With respect to TCE, two isotope ratios can be confidently distinguishable if the discrepancy between the isotope ratios is greater than 0.0021. If the highest precisions (0.21‰ for PCE, 0.25‰ for TCE) are reached, then the thresholds of isotope-ratio differences at 0.0008 and 0.0010 can be clearly distinguishable for PCE and TCE, respectively.

The standard deviations of relative isotope-ratio variations ($\delta^{37}$Cl') of PCE were 0.64-1.22‰, which were relatively lower than those of TCE (0.80-1.69‰). Therefore, if the $\delta^{37}$Cl' is higher than 4.9‰ for PCE or larger than 6.8‰ for TCE, then these isotope ratios can be confidently differentiated. If the lowest $\delta^{37}$Cl' standard deviations (0.64‰ for PCE, 0.80‰ for TCE) are available, the critical $\delta^{37}$Cl' for differentiating two isotope ratios are 2.6‰ and 3.2‰ for PCE and TCE, respectively. This demonstrates that the highest precisions of the present methods are comparable to those in a previous study using GC-qMS, in which the distinguishable thresholds of $\delta^{37}$Cl' were 2.0-4.0‰ for TCE depending on different types of GC-qMS.[18]

*Linearity and Sensitivity.* Two concentration levels (1.0 μg/mL and 0.1 μg/mL) were tested for investigating the mass dependency of chlorine isotope ratio analysis. As documented in Table S-2, the standard deviations increased by one or two folds as the concentrations from 1.0 μg/mL to 0.1 μg/mL, which indicated that the precisions were concentration-dependent. While the measured isotope ratios showed no statistically significant drift as the concentrations varied. The isotope-ratio standard deviation of PCE at 0.1 μg/mL was 0.49‰, which may be somewhat competent in CSIA-Cl when isotope ratio differences are relatively large. However, the isotope-ratio standard deviation of TCE at 0.1 μg/mL (0.98‰) may not be capable of differentiating isotope ratios in most cases in the real world. Therefore, the lower limit of detection of PCE was 0.1 μg/mL, and that for TCE was proposed to be 1.0 μg/mL. The injected amount of the lower limit of detection of PCE was 0.1 ng, which is lower than that in a reported study using GC-qMS[15] and at least one order of magnitude lower than that using GC-IRMS,[9] and significantly lower than the analysis amounts for $\delta^{37}$Cl analysis using off-line TIMS (3 μg of chloride)[25] and GC-high-temperature conversion-MS/IRMS (15 nmol Cl).[13]

**Evaluation of Different Calculation Schemes of Isotope Ratio.**

So far, the most used isotope-ratio calculation schemes for CSIA-Cl/Br using GC-EI-MS are on the basis of the binomial theorem. These relevant studies tacitly approved that the isotopologues of the target organochlorines and the isotopologue ions of these compounds on GC-EI-MS conform to binomial distribution. The isotope ratio thus can be calculated with any pair of adjacent isotopologues of either molecular ions or product ions via the isotopologue-pair scheme:[20]

$$IR = \frac{i}{n-i+1} \times \frac{I_i}{I_{i-1}} \quad (4)$$

where $n$ is the number of chlorine atoms of a certain ion; $i$ is the number of $^{37}Cl$ atoms in a specific isotopologue ion; and $I$ denotes the MS signal intensities.

Besides the complete-isotopologue scheme of isotope ratio calculation, we also used the isotopologue-pair scheme to calculate the chlorine isotope ratios of PCE and TCE for comparing the performances of the two schemes. As indicated in Table 1, the standard deviations of isotope ratios calculated with the isotopologue-pair scheme were 1.2-2.9 folds of the corresponding values calculated with the complete-isotopologue scheme, which demonstrated significantly higher precisions of complete-isotopologue scheme than the isotopologue-pair scheme. The standard deviations of the $δ^{37}Cl'$ obtained with the two schemes also showed the similar scenario (Table 1). Moreover, as shown in Figure 1 and Table 1, the isotope ratios calculated with the complete-isotopologue scheme were always lower than those calculated with the isotopologue-pair scheme, particularly for the PCE standard from manufacturer-1, of which the isotope ratio discrepancies were up to 11.1-11.5‰. The differences of isotope ratios calculated with the two schemes of the TCE standard from manufacturer-1 and the PCE standard from manufacturer-2 were smaller than those of the PCE standard from manufacturer-1, but still with statistical significance (2.6-4.8‰). To the contrary, isotope ratios calculated with the different schemes of the TCE standard from manufacturer-2 exhibited no statistically significant difference (<0.2‰). These results indicate that the molecular-ion isotopologues of the standards of PCE and TCE on the GC-HRMS did not comply with binomial distribution, which has been theoretically proved in our previous study.[23] Furthermore, as theoretically proved in our previous study, the isotopologues of the synthetic standards of PCE and TCE might not be binomially distributed, and isotopologues of environmental organochlorines are extremely impossible to follow binomial distribution.[24]

In addition, as illustrated in Figure 1, the variation tendencies of isotope ratios of the standards from the two manufacturers calculated with different schemes were inconsistent. This resulted in obviously contradictory $δ^{37}Cl'$ obtained with different schemes when taking a standard form a manufacturer as the external isotopic standard for the corresponding standard from another manufacturer. As Table 1 shows, the $δ^{37}Cl'$ of the standards calculated with the complete-isotopologue scheme were completely contradictory to the corresponding $δ^{37}Cl'$ calculated with the isotopologue-pair scheme. Therefore, to our point of view, the complete-isotopologue scheme rather than the isotopologue-pair scheme should be proposed for CSIA-Cl when detection is performed by GC-EI-MS. When using the isotopologue-pair scheme, analysts should take account of the specific isotopologue distributions of external isotopic standards and analytes, which may trigger biased results in CSIA-Cl.

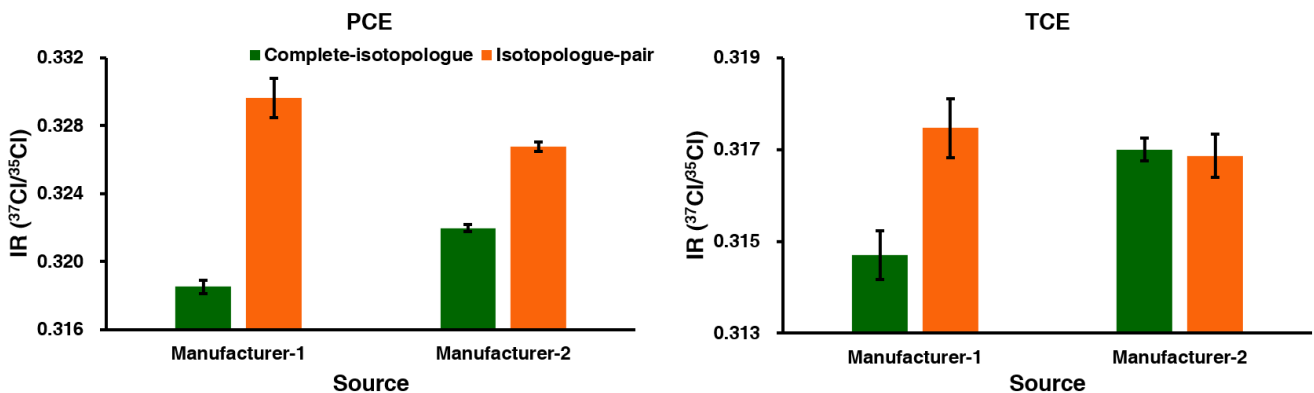

**Figure 1.** Isotope ratios of perchlorethylene (PCE) and trichloroethylene (TCE) standards from different manufacturers calculated with different isotope ratio evaluation schemes. Error bars denote the standard deviations (1σ, n=5). IR: isotope ratio.

**Impacts of Isotope Fractionation Occurring on GC-MS.**

As found in our previous study, many organochlorines can exhibit chlorine fractionation on GC columns, which may lead to deviations if rational integration of chromatographic peaks is not achieved.[21] In the present study, PCE and TCE were found to show no observable chlorine isotope fractionation on the GC column (Figure 2a-b). Therefore, the on-column isotope fractionation was not a problematic issue needing consideration. In a previous study, we found that many organochlorines showed chorine isotope fractionation in EI-MS, which was generally much more significant than that occurring on GC columns.[22] In this work, we also investigated the chlorine isotope fractionation of the analytes during fragmentation in EI-MS. As shown in Figure 2c-d, significant isotope fractionation was found for both PCE and TCE in EI-MS, particularly between the P-Cl$_2$ (product ion with 2 Cl atoms) ion and the P-Cl$_1$ ion, of which the isotope-ratio differences were 0.0533±0.0012 and 0.0566±0.0005 for PCE and TCE, respectively. Therefore, the isotope ratios derived from different ions are unequal, which contradicts the conclusions in some previous studies.[3,16,20]

Moreover, as shown in Figure 2c-d and Table S-3, the isotope ratios calculated with the molecular-ion isotopologues generally presented higher precisions than those calculated with the product ions. This might attribute to that the molecular ions were merely influenced by intermolecular fractionation, while the product ions might be affected by both intermolecular and intramolecular fractionations, of which the latter are always more significant than the former.[26] More significant isotope fractionation during fragmentation in EI-MS may lead to higher uncertainties in measurement of isotope ratios. Therefore, molecular ions are recommended to perform isotope ratio calculation in CSIA-Cl for achieving high-quality data.

Since the chorine isotope fractionation in EI-MS is very stable (Figure 2c-d), then it may not able to impact the precision and trueness of CSIA-Cl if external isotopic standards are employed.

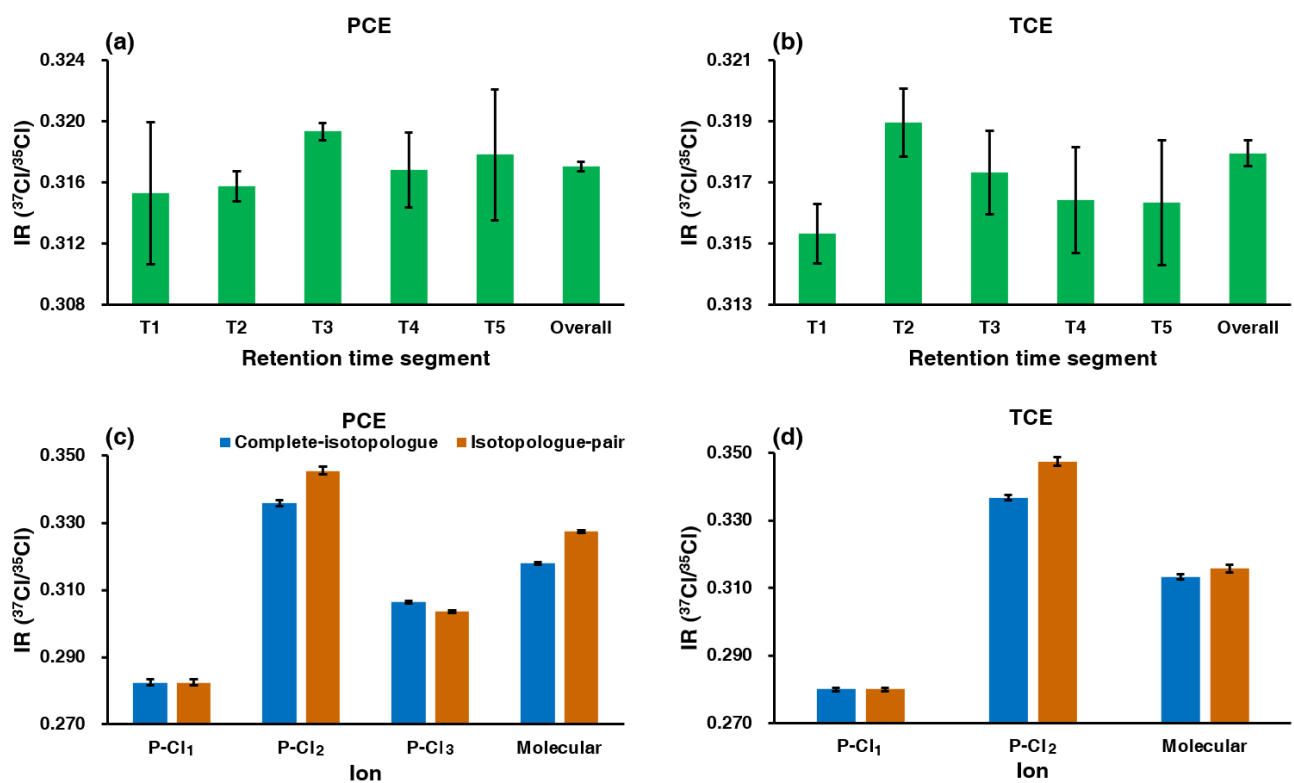

**Figure 2.** Chlorine isotope fractionations of PCE and TCE occurring on the gas chromatographic column and in the electron ionization mass spectrometer. (**a**): the isotope ratios of PCE in different retention-time segments (T1-T5) and in overall retention-time range (overall), (**b**): the isotope ratios of TCE in different retention-time segments and in overall retention-time range, (**c**): the isotope ratios of PCE derived from the different product ions (P-Cl$_1$ to P-Cl$_3$) and the molecular ion with different isotope-ratio calculation schemes; (**d**): the isotope ratios of TCE derived from the different product ions (P-Cl$_1$ and P-Cl$_2$) and the molecular ion with different isotope-ratio calculation schemes. Error bars denote the standard deviations (1σ, n=5). T1-T5: the retention-time segments 1 to 5; P-Cl$_x$: product ions with $x$ Cl atoms.

**Temporal Isotope Ratio Drifts with Injection Sequences.**

As indicated in Figure 3, the isotope ratios of all the standards in individual analysis batch are relatively stable. The PCE standard from manufacturer-1 showed random fluctuations of isotope ratios around the average. However, the PCE standard from manufacturer-2 and the TCE standards from

manufacturer-1 and manufacturer-2 showed the isotope-ratio drifts of 0.8‰, 3.1‰ and 3.1‰, respectively. This result demonstrates that the application of external isotopic standards is mandatory in CSIA-Cl. In addition, too many bracketing injection replicates and overlong time of an analysis batch for an analyte should be avoided in order to obtain satisfactory data.

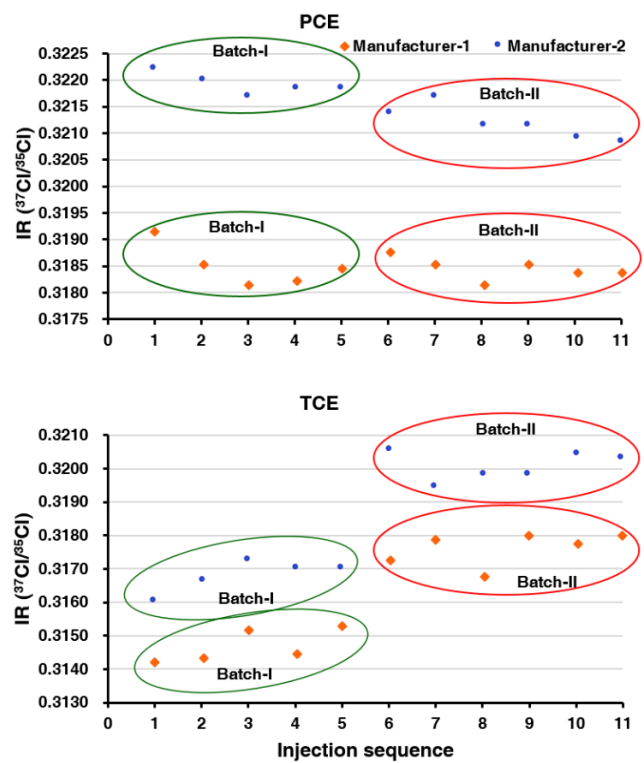

**Figure 3**. Temporal drifts of isotope ratios with injection sequences of the standards of PCE and TCE. The interval between the first and the second batches was one day.

**Comparison with Previous CISA-Cl Methods.**

Several studies have reported the CSIA-Cl methods using GC-qMS[14-17] or GC-QTOF-MS.[19] All these methods used the isotopologue-pair scheme to calculate the isotope ratios. As elucidated above, the isotopologue-pair scheme may lead to biased data if the target chlorinated compounds in the sample have specific chlorine isotopologue distributions in comparison with the corresponding external isotopic standards. In addition, the isotopologue distributions of detected ions of organochlorines in EI-MS by no means are binomial. Therefore, the isotopologue-pair scheme, based on the prerequisite assumption that the chlorine isotopologues of detected ions of analytes are binomial distributed, are not procedurally correct, even though the achieved data could reflect the trueness in some cases.[3,15,18] This scenario may be an example of near miss which means that a scheme can approach some correct conclusions or results but it is wrong indeed.[27] The complete-isotopologue scheme and the previously reported complete-ion method can avoid these drawbacks for CSIA-Cl.[16] Unfortunately, the complete-isotopologue scheme and the complete-ion method may not suitable for CSIA-Cl of organochlorines with more than two chlorine atoms when using GC-qMS form detection, due to insufficient sensitivity of this instrument.[16] However, GC-DFS-HRMS can provide sufficient signal intensities for organochlorines with the number of Cl atoms up to at least four for CSIA-Cl using the complete-isotopologue scheme with fairly low injection amounts (around 1 ng). In addition, the high selectivity of GC-DFS-HRMS can assure the measurement of chlorine isotopologue ions free from possible interference which may be a problem in analysis using GC-qMS due to its low mass resolution. On the other hand, GC-DFS-HRMS may be with low accessibility, high cost and difficulty of operation, which are the disadvantages in comparison with widely used bench-top GC-qMS.

**CONCLUSIONS**

CSIA-Cl is a practicable and high-efficient measure to quantify transformation processes and identify pollution sources for chlorinated organic compounds, and on the way to improvement yet. In this study, we developed a CSIA-Cl method for PCE and TCE using GC-DFS-HRMS through injection of standards in a bracketing way along with the complete-isotopologue scheme of isotope ratio calculation using molecular ions merely. The obtained highest precisions (standard deviations of isotope ratios) were 0.21‰ and 0.25‰ for PCE and TCE, respectively. On the other hand, the achieved lowest standard deviations of relative isotope ratio variations ($\delta^{37}Cl'$) between two corresponding standards of the same analyte from different manufacturers were 0.64‰ and 0.80‰ for PCE and TCE, respectively. As a consequence, the thresholds of $\delta^{37}Cl'$ for distinguishing two isotope ratios were 2.6‰ and 3.2‰ for PCE and TCE, respectively based on the developed method, which were comparable with those of some reported analogous methods using GC-qMS. In addition, the lower limits of detection for CSIA-Cl of the analytes were lower than those of most previous studies. Two schemes of isotope ratio calculation, namely, complete-isotopologue scheme and isotopologue-pair scheme, were compared in terms of precision and accuracy. The complete-isotopologue scheme is superior to the isotopologue-pair scheme with respect to precision and reflection of trueness. The chlorine isotope fractionation taking place on GC-MS was found to trigger no significantly negative impact on the precision and accuracy of CSIA-Cl if external isotopic standards were applied to calibration. We proposed that external isotopic standards should be applied and injected along with the analysis of target samples in CSIA-Cl using GC-HRMS or GC-qMS, in the light of the observation of the isotope-ratio drifts with injection sequences of some standards in this study. This method will help to conduct the future studies involving quantification of transformation processes and source apportionment of PCE and TCE, and can be expanded to method development of CSIA-Cl for a large number of other chlorinated organic compounds.

## ASSOCIATED CONTENT

The Supporting Information is available free of charge on the website at http://pending.

## ACKNOWLEDGEMENTS

This work was partially financed by the National Natural Science Foundation of China (Grant No. 41603092).

## AUTHOR INFORMATION

**Corresponding Author**

*E-mail: CaimingTang@gig.ac.cn (C. Tang).

**Notes**

The authors declare no competing financial interest.

**Table**

**Table 1.** Measured raw isotope ratios and relative isotope ratio variations ($\delta^{37}Cl'$) of perchlorethylene (PCE) and trichloroethylene (TCE) from different manufacturers with two different isotope ratio calculation schemes (the complete-isotopologue scheme and the isotopologue-pair scheme).

| Calculation scheme | Batch | Compound | Manufacturer-1 | | Manufacturer-2 | | Manufacturer-1 vs manufacturer-2 | | Manufacturer-2 vs manufacturer-1 | |
|---|---|---|---|---|---|---|---|---|---|---|
| | | | Mean isotope ratio | Standard deviation (1σ, ‰) | Mean isotope ratio | Standard deviation (1σ, ‰) | $\delta^{37}Cl'$ (‰) | Standard deviation (1σ, ‰) | $\delta^{37}Cl'$ (‰) | Standard deviation (1σ, ‰) |
| Complete-isotopologue scheme | Batch-I (n=5)[a] | PCE | 0.31851 | 0.39 | 0.32197 | 0.21 | -10.74 | 1.22 | 11.36 | 0.66 |
| | | TCE | 0.31470 | 0.53 | 0.31700 | 0.25 | -7.24 | 1.69 | 7.29 | 0.80 |
| | Batch-II (n=6) | PCE | 0.31848 | 0.21 | 0.32121 | 0.29 | -8.72 | 0.64 | 8.58 | 0.92 |
| | | TCE | 0.31776 | 0.33 | 0.32006 | 0.43 | -7.19 | 1.04 | 7.24 | 1.37 |
| Isotopologue-pair scheme | Batch-I (n=5) | PCE | 0.32961 | 1.14 | 0.32675 | 0.28 | 8.77 | 3.50 | -8.69 | 0.86 |
| | | TCE | 0.31747 | 0.64 | 0.31686 | 0.47 | 1.92 | 2.03 | -1.91 | 1.48 |
| | Batch-II (n=6) | PCE | 0.32994 | 0.40 | 0.32603 | 0.54 | 12.00 | 1.21 | -12.11 | 1.65 |
| | | TCE | 0.32035 | 0.60 | 0.31994 | 0.51 | 1.28 | 1.87 | -1.27 | 1.61 |

Note, a: number of injection replicates.

# Table of Contents

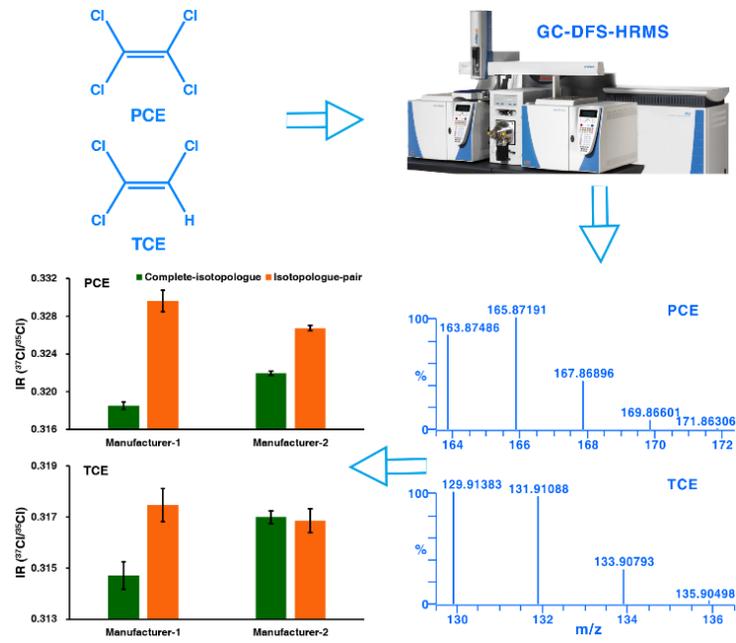